\documentclass[aps,prb,twocolumn,amsmath,amssymb,fleqn,superscriptaddress]{revtex4-2}
\usepackage{graphicx}
\usepackage{bm}
\usepackage{subfigure}
\usepackage{booktabs}
\usepackage{color}

\usepackage[usenames,dvipsnames,svgnames,table]{xcolor}
\definecolor{forestgreen}{rgb}{0.11,0.54,0.15}
\definecolor{purple}{rgb}{0.62,0.10,0.96}
\definecolor{dockerblue}{rgb}{0.11,0.56,0.98}
\definecolor{freeblue}{rgb}{0.25,0.41,0.88}
\usepackage[pdftex,plainpages=false,colorlinks=true,linkcolor=Red, citecolor=blue, urlcolor=blue]{hyperref}

\begin{document}  

\title{Disorder-enhanced effective masses and deviations from Matthiessen's rule in PdCoO$_2$ thin films}

\author{David Barbalas}\email{dbarbalas@jhu.edu}
\affiliation{Department of Physics and Astronomy,
The Johns Hopkins University, Baltimore, Maryland 21218, USA}

\author{Ana\"{e}lle Legros}
\affiliation{Department of Physics and Astronomy,
The Johns Hopkins University, Baltimore, Maryland 21218, USA}

\author{Gaurab Rimal}
\affiliation{Department of Physics and Astronomy, Rutgers, The State University of New Jersey,
Piscataway, New Jersey 08854, US}

\author{Seongshik Oh}
\affiliation{Department of Physics and Astronomy, Rutgers, The State University of New Jersey,
Piscataway, New Jersey 08854, US}

\author{N. P. Armitage}\email{npa@jhu.edu}
\affiliation{Department of Physics and Astronomy,
The Johns Hopkins University, Baltimore, Maryland 21218, USA}

\date{\today}

\begin{abstract}

The observation of hydrodynamic transport in  the metallic delafossite PdCoO$_2$ has increased interest in this family of highly conductive oxides, but experimental studies so far have mostly been confined to bulk crystals. In this work, the development of high-quality epitaxial thin films of PdCoO$_2$ has enabled a thorough study of the conductivity as a function of film thickness using both dc transport and time-domain THz spectroscopy.  As film thickness increases from 12 nm to 102 nm, the residual resistivity decreases and we observe a large deviation from Matthiessen's rule (DMR) in the temperature dependence of the resistivity.  We find that the complex THz conductivity is well fit by a single Drude term.  We fit the data to extract the spectral weight and scattering rate simultaneously. The temperature dependence of the Drude scattering rate is found to be nearly independent of the residual resistivity and cannot be the primary mechanism for the observed DMR. Rather, we observe large changes in the spectral weight as a function of disorder, changing by a factor of 1.5 from the most disordered to least disordered films.  We believe this corresponds to a mass enhancement of $\geq 2$ times the value of the bulk effective mass which increases with residual disorder. This suggests that the mechanism behind the DMR observed in dc resistivity is primarily driven by changes in the electron mass. We discuss the possible origins of this behavior including the possibility of disorder-enhanced electron-phonon scattering, which can be systematically tuned by film thickness. 
\end{abstract}

\maketitle

\section{Introduction}
Since their initial discovery, the metallic delafossites have been noted for their extremely high in-plane conductivity \cite{shannon_chemistry_1971,rogers_chemistry_1971}.  Developments in the growth of high quality single crystals have shown that the room temperature resistivity is comparable to that of the noble metals \cite{takatsu_roles_2007,hicks_quantum_2012}. The high conductivity in PdCoO$_2$ is notable as it originates from the highly anisotropic transport between the ab plane and along the c-axis ($\rho_{ab}/\rho_c \geq 100$) \cite{hicks_quantum_2012,daou_large_2015} due to the separation of the metallic Pd-layers and the insulating CoO$_2$ octahedra layers. This compound possesses the largest mean free path for oxides at low temperature, $l \sim $ 20 $\mu$m \cite{hicks_quantum_2012}, which is an extremely large value even as compared to the noble metals. Signatures of a hydrodynamic response  \cite{moll_evidence_2016,scaffidi_hydrodynamic_2017,nandi_unconventional_2018} and recent work studying phenomena originating from the hexagonal Fermi surface \cite{cook_electron_2019} and the quasi-2D nature of the conductivity has stimulated the growing interest in this compound \cite{putzke_he_2020,harada_thin-film_2021}.

\begin{figure*}[!ht]
    \centering
    \includegraphics[width = 1\textwidth]{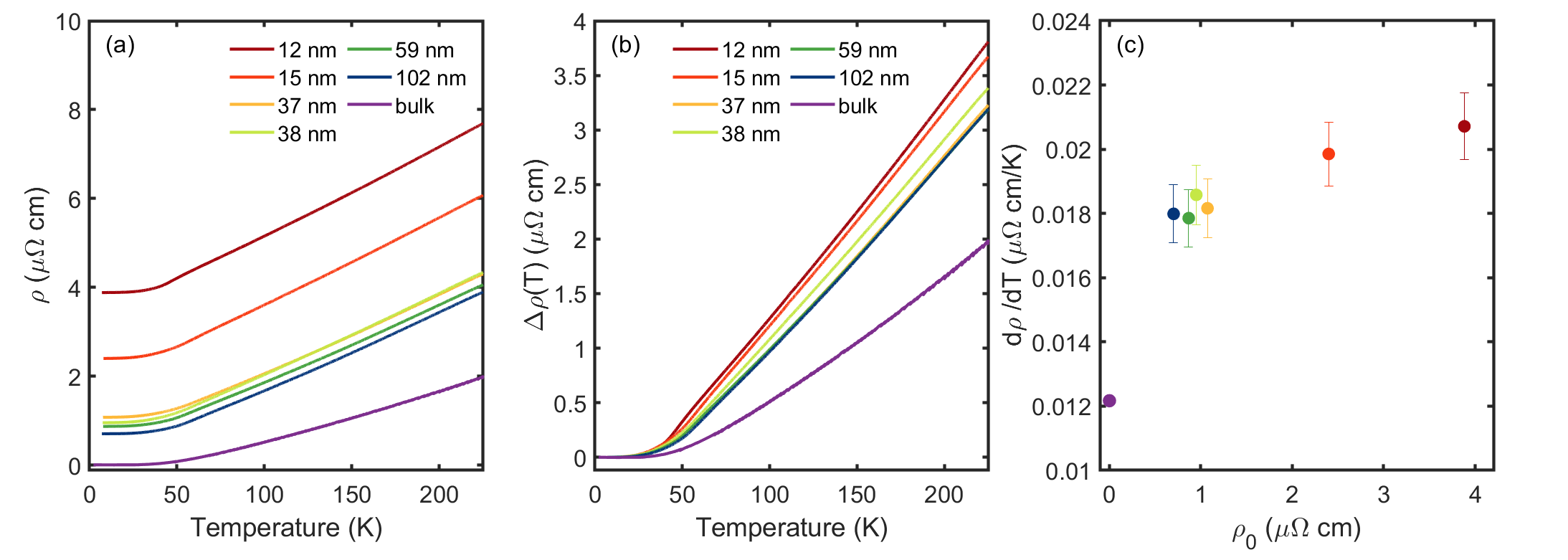}
    \caption{(a) The  dc resistivity (a) shown for all samples studied, along with data for bulk crystals (RRR $\sim 330$) from Ref.~\cite{li_situ_2019} for comparison. (b) The relative temperature dependence $\Delta\rho(T) = \rho(T) - \rho_0$, where $\rho_0 = \rho(5$ K$)$. Note how as film thickness decreases, the temperature dependent part of the resistivity grows in magnitude. It should also be noted that the temperature dependence of the films studied at low temperature do not display a large deviation from the behavior from the bulk crystal. (c) The high temperature slope of the resistivity $ d\rho/dT$ used to parameterize the deviation from Matthiessen's rule (DMR) observed in the temperature dependence of the resistivity above 100 K. The films demonstrate a sizeable increase in the high temperature slope compared to the single crystal. The slope of the resistivity in the thin films is an increasing function of the residual resistivity, indicating a positive DMR.}
    \label{DCResistivity}
\end{figure*}

Before the development of thin films, the high conductivity of single crystals restricted optical studies of PdCoO$_2$ using a reflection geometry \cite{homes_perfect_2019}; however, this made studying the low-energy response extremely difficult as the reflectivity of the material is nearly indistinguishable from a perfect mirror. The first development of thin films of PdCoO$_2$ via PLD \cite{harada_highly_2018} has lead to subsequent work on thin film deposition \cite{yordanov_large_2019,sun_growth_2019,brahlek_growth_2019, wei_solution-processable_2020,harada_thin-film_2021} which presents new opportunities to study intrinsic material properties with new experimental techniques \cite{mackenzie_properties_2017}. With the development of high quality epitaxial thin films, optical measurements in the transmission geometry are now possible. This presents previously unexplored avenues for studying both the linear response and nonlinear optical response, which is predicted to host unique signatures in the bulk hydrodynamic regime \cite{forcella_electromagnetic_2014,sun_universal_2018}. Optical techniques are well suited to studying metallic systems in linear response as they are non-contact probes that can provide information about both the quasiparticle scattering rates and Drude spectral weight simultaneously. Nonlinear THz spectroscopy is also predicted to be able to observe hydrodynamic signatures of coupled electron-electron or electron-phonon fluids \cite{sun_universal_2018,levchenko_transport_2020,huang_electron-phonon_2021} due to unique frequency-dependent features in the optical response. 

While the majority of experiments on transport in PdCoO$_2$ have focused on the low disorder limit in ultra-clean single crystals \cite{nandi_unconventional_2018,putzke_he_2020,bachmann_super-geometric_2019}, in thin film samples the film thickness and annealing process can be used to tune samples across a large range of disorder in a repeatable fashion. While the intrinsic quasi-2D nature of the in-plane conductivity and the long mean free paths of carriers have been well characterized in the ultra-clean limit, the role of disorder and its influence on transport have only been systematically studied in the sister compound PtCoO$_2$ as a function of electron irradiation \cite{sunko_controlled_2020}. Elucidating what effect higher disorder levels in PdCoO$_2$ thin films have on the conductivity is crucial to understand the nature of in-plane transport and for future applications in heterostructures and devices. 

Studying the role of disorder in metals is critical to understand the relevant scattering mechanisms, where tuning disorder can reveal possible correlations between different scattering channels in metals. Generically, Matthiessen's rule assumes that if the relevant scattering mechanisms are independent, the total scattering rate will be given as the sum $1/\tau = 1/\tau_1 + 1/\tau_2 + ...$. Deviations from Matthiessen's rule (DMR) can highlight departures from an ideal free electron gas such as the presence of anisotropic scattering. For example, in simple metals such as Cu, Ag or Au, all show positive DMR due to anisotropy of the phonon-electron interaction between the spherical and neck segments of the Fermi surface \cite{dugdale_mathiessens_1967}. The nature of the specific disorder introduced (dislocations, strain, neutral vs. charged impurities) all have distinct characteristics that can also be inferred from the temperature dependence of the observed DMR. 

There also exist interesting phenomena in the limit of strong disorder in metals, such as disorder-driven metal-insulator transitions where the insulating state is driven by localization \cite{anderson_absence_1958}. Within correlated metallic systems, the importance of disorder and its interactions has become more prominent recently, but deserves more thorough study \cite{green_quantum_2018, vojta_disorder_2019}. The role of disorder in modifying the nature of transport in metallic systems can be used to determine both the origin of scattering in the clean limit as well as play a role in enhancing existing interactions \cite{bass_deviations_1972}. The tunability of thin films also provides an opportunity to determine if features found in bulk PdCoO$_2$ samples including inter-layer effects and long in-plane coherent lengths are robust in thin film devices or heterostructures  \cite{lu_layer-resolved_2021,yim_quasiparticle_2021}. 

In this work, we combine 4-probe dc resistivity measurements and time-domain THz spectroscopy (TDTS) to study the low-frequency optical conductivity of PdCoO$_2$ thin films. We studied thin films with thicknesses from 12 - 102 nm, where the residual resistivity $\rho_0$ changes by more than a factor of 5$\times$ from the thickest film (0.7 $\mu\Omega$ cm) to the thinnest film (3.9 $\mu\Omega$ cm). We find that the temperature dependence of the resistivity is comparable to single crystals for the thicker films. The dc resistivity on these films shows a sizeable positive deviation from Matthiessen's rule (DMR), where the temperature dependence of the resistivity increases as the film thickness decreases. In order to further study the mechanisms driving the DMR, we fit the complex conductivity from TDTS to a single Drude term to extract the intraband scattering rate and the Drude spectral weight (which is proportional to the ratio of the carrier density to the mass). We find that the scattering rate is nearly independent of the residual disorder in the films, but the Drude spectral weight appears to decrease by a factor of 1.5$\times$ as the disorder is increased. From Hall effect measurements on the films, the 3D carrier density $n_{3D}$ at the lowest temperatures is inferred to be independent of film thickness, suggesting that the changes in the spectral weight are due to renormalization effects encapsulated in an effective mass enhancement of the quasiparticles that increases with residual disorder. In addition to making the THz transmission experiments possible, our work also shows that the development of high quality thin films of PdCoO$_2$ are a robust platform that retain the highly conductive properties of bulk crystals while also demonstrating the ease of tunability suitable for studying the effects of disorder.

\section{Methods}
The single crystalline thin films of PdCoO$_2$ used in this study had thicknesses in the range of 12-102 nm and were grown on (0001) sapphire substrates using molecular-beam epitaxy (MBE). The films were characterized by x-ray photoemission spectroscopy (XPS) and x-ray diffraction (XRD) to ensure high quality thin films. The thickness of the films was measured using Rutherford backscattering (RBS) to a precision of $3-5\%$ . The dc resistivity was conducted in the van der Pauw geometry to measure the (ab) plane transport, and Hall effect measurements were conducted to determine the carrier density $n_{3D}$. More details on the film growth can be found elsewhere \cite{brahlek_growth_2019}. The thin films studied had residual resistivity ratios (RRR = $\rho_{250K}/\rho_{5K}$) up to 6; for thicker films with $t \geq 150$ nm the RRR increases up to 10. The material parameters of the films used for this study are given in Table.~\ref{table} We observe that the residual resistivity ratio is sensitive to the film thickness, as thicker films have a smaller proportion of surface scattering as well as a reduced concentration of in-plane defects.

\begin{table}[b]
\begin{center}
\begin{tabular}{|c|c|c|c|c|}
\hline
Film Thickness  & $\rho(250K)$ & $\rho(5K)$ & RRR  & RRR \\
(nm) & ($\mu\Omega \; \mathrm{cm}$) & ($\mu\Omega \; \mathrm{cm}$) & (dc) & (THz)\\
\hline
\hline
12  &  8.24 &  3.88 &  2.1 & 2.4 \\
15 & 6.56  & 2.40  & 2.7  & 3.0 \\
37 &  4.8 & 1.07  &  4.5  & 5.4\\
 38 &  4.75  & 0.95  &  5.0  & 5.5 \\
 59 &  4.52   & 0.87 &  5.2  & 6.7\\
102  & 4.34    & 0.70  &  6.2  & 10.3\\
\hline
\end{tabular}
\end{center}
\caption{Samples used for the study. Thicknesses range from 12 nm to 102 nm with error of $\pm 0.5$ nm. The residual resistivity ratio (RRR) here is defined as the ratio of the resistivity at 250 K and 5 K. The RRR is shown both from the dc resistivity measurement, as well as from extrapolation from the THz conductivity $\sigma(\omega \rightarrow 0)$. For reference, high quality single crystals have RRR $\geq 100-300$ \cite{tanaka_origin_1998,hicks_quantum_2012}. }
\label{table}
\end{table}

The low-frequency response of the films were characterized using time-domain THz spectroscopy (TDTS) allowing for accurate determination of both the real and imaginary parts of the frequency dependent conductivity from the THz complex transmission.  The complex transmission was obtained by referencing the transmitted signal through the sample and comparing to a bare sapphire substrate. From the complex transmission, the complex conductivity can be directly determined in the thin film approximation as
\begin{equation}
	\widetilde{T}(\omega)= \frac{n_s + 1}{n_s + 1 + Z_0 d \: \widetilde{\sigma}(\omega)} e^{i  \frac{\omega}{c}\Delta L (n_s - 1)}   ,
\end{equation}
where $n_s$ is the substrate index of refraction, $Z_0$ is the impedance of free space, $d$ is the sample thickness, $\widetilde{\sigma}$ is the complex conductivity, $\omega$ is the measurement frequency, $c$ is the speed of light, and $\Delta L$ is a correction factor on the order of $\sim 1 \;\mu$m to account for differences in thickness between the substrate and the sample calculated using the numerical method \cite{krewer_accurate_2018}. Uncertainty in $\Delta L$ affects the extracted phase of the conductivity and hence the relative values of $\sigma_1$ and $\sigma_2$; the use of the aforementioned numerical method allows for the $\Delta L$ to be determined within $\pm 0.5$ $\mu$m. Due to the high conductivity and strong frequency dependence of the THz transmission, it is the largest source of error in extracting fitting parameters of the conductivity as discussed below.

The advantage of TDTS as a non-contact probe in studying metallic systems is that it can directly obtain the frequency dependent optical conductivity. A simple but useful model for understanding the intraband conductivity of metals is the Drude model given as
\begin{equation}
    \sigma(\omega) = \sigma_1 + i \sigma_2 =  \frac{\omega_p^2 \tau}{1+\omega^2\tau^2} + i \frac{(\omega_p^2 \tau)\omega\tau}{1+\omega^2\tau^2}
\end{equation}
where $\omega_p = \sqrt{\frac{ne^2}{m}}$ is the plasma frequency and $\tau$ is the scattering time. Here $n$ is the charge density and $m$ is a transport mass. From simultaneous fitting of the real and imaginary conductivity, the intraband scattering rate (given by the width of the Lorentzian Drude peak in frequency $\Gamma = 1/2\pi \tau$) and spectral weight $\omega_p^2$ can be directly obtained.  Note that the real and imaginary parts of a response function are related to each other via a Kramers-Kronig relation. 

\begin{figure*}[!ht]
    \begin{center}
    	\includegraphics[width= \textwidth]{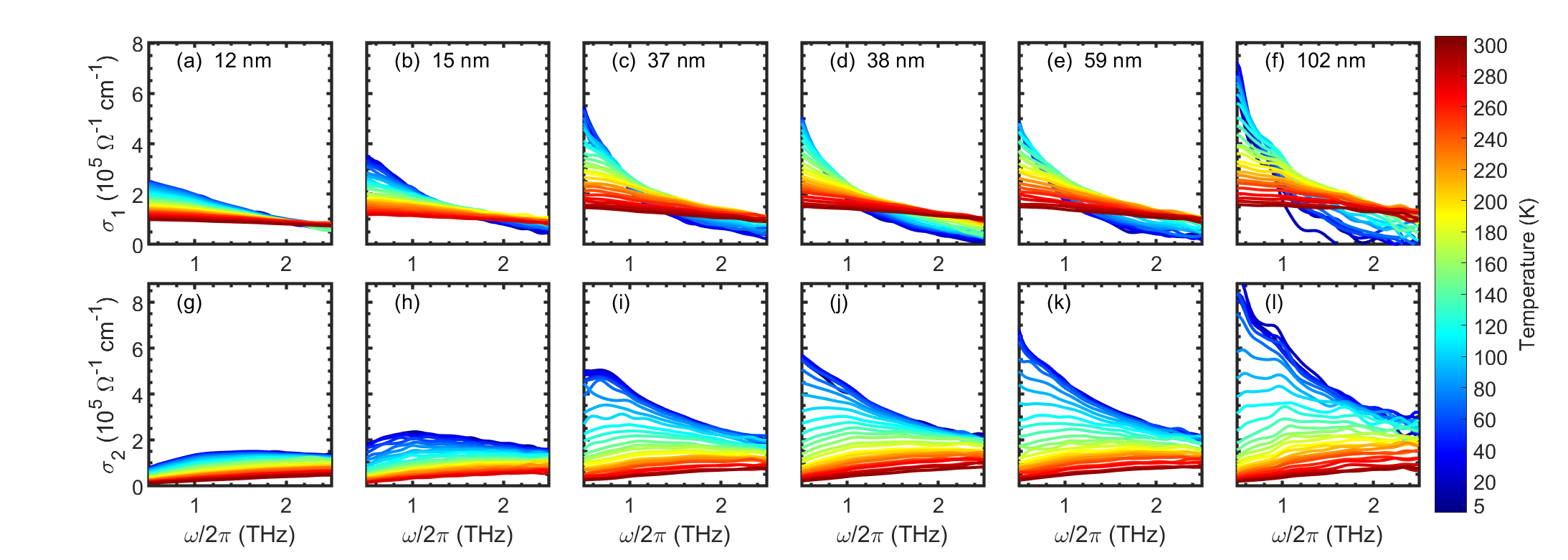}
        \caption{Real (a-f) and imaginary (g-l) THz conductivity for all six PdCoO$_2$ films studied starting with the thinnest 12 nm film on the left. Note that upon increasing the film thickness, the conductivity rapidly increases. The conductivity shows good agreement with the Drude model and upon cooling the scattering rate can be seen to move from above our spectral range to below for the cleaner samples (as defined by the Drude model as the crossing point of $\sigma_1$ and $\sigma_2$) and the maximum in  $\sigma_2$. }
        \label{sigma_all}
    \end{center} 
\end{figure*}

\begin{figure}[h]
    \centering
    \includegraphics[width=1 \columnwidth]{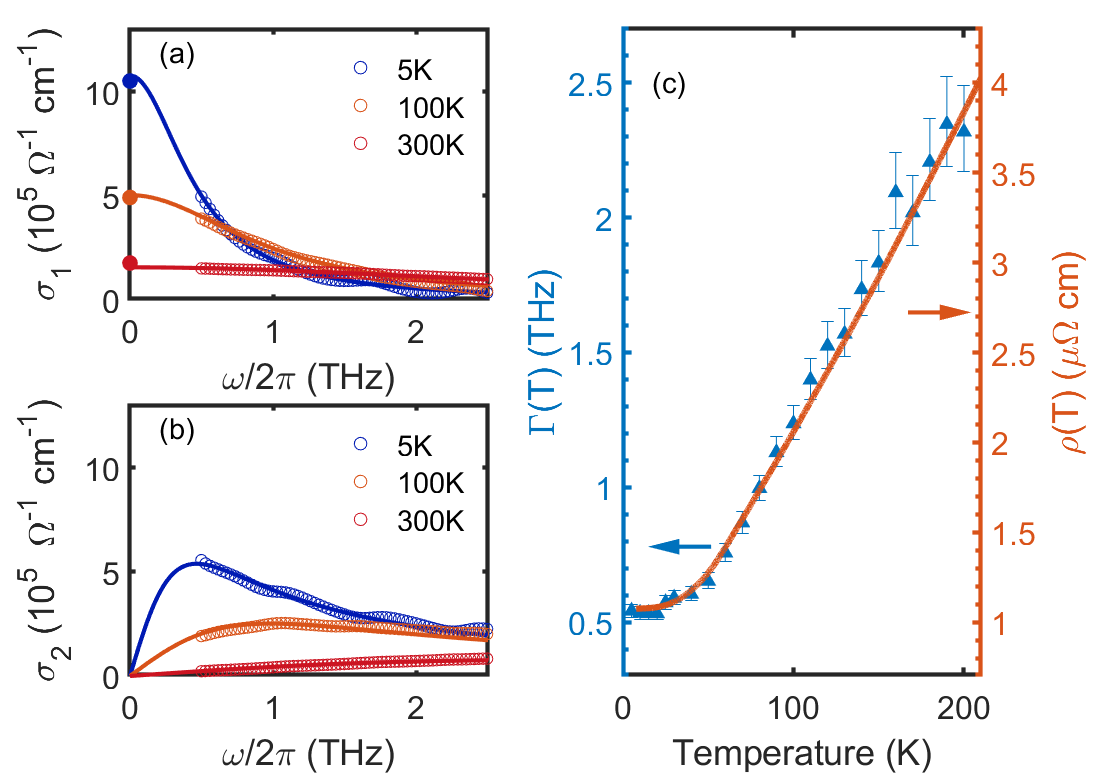}
    \caption{The real (a) and imaginary (b) conductivity for the 37 nm PdCoO$_2$ film with Drude fits shown for selected temperatures along with the dc conductivity value. Comparison of the resistivity and the scattering rate are shown in (c); here the error bars are given by the change in fit parameters due to changes in $\Delta L $ of $\pm 0.5 \;\mu$m in the complex conductivity calculation from the THz transmission. The error in the Drude fits is negligible.}
    \label{sigma_35nm}
\end{figure}

\section{Experimental Results}
In Fig.~\ref{DCResistivity}(a), the dc resistivity for the films is shown alongside the resistivity of a high quality bulk crystal \cite{li_situ_2019}. By decreasing the thickness of the film, the residual resistivity increases; the films also demonstrate an increase in magnitude of temperature dependent component of the resistivity. At low temperatures, the mobility is measured to be As a function of thickness, the films demonstrate a positive deviation from Matthiessen's rule (DMR) seen as a systematic increase in the temperature dependent component of resistivity as a function of increasing disorder. This trend is more clearly seen in Fig.~\ref{DCResistivity}(b) where only the temperature dependent part of the resistivity  $\Delta\rho(T) = \rho(T) - \rho_0$ is shown for all films.  

Naively, one would expect that if Matthiessen's rule was obeyed, the temperature dependence of the resistivity should be independent of the residual resistivity. In the simplest picture of Matthiessens' rule, scattering channels are additive and that the temperature dependent inelastic scattering  (electron-phonon, Umklapp electron-electron, etc.) should contribute independently of elastic scattering from disorder. However, it is clear from Fig.~\ref{DCResistivity}(b) that the temperature dependent part of the dc resistivity for the thin films demonstrates clear changes in the high temperature slope. This indicates that the various contributions to the resistivity are not fully independent of each other.  While DMR is expected to be present in all materials due to the anisotropy of relaxation times for phonon and impurity scattering \cite{dugdale_mathiessens_1967,bass_deviations_1972}, the size of the deviations are typically small as compared to $\rho(T)$ or $\rho_{imp}$. However, the magnitude of DMR observed in these PdCoO$_2$ thin films is significantly larger (3 $\mu\Omega$ cm at $T = 300$ K) than what has been seen by controlled electron irradiation ($< 0.5 \,\mu\Omega \,$cm at $T = 300$ K) in single crystals of PdCoO$_2$ and PtCoO2 \cite{sunko_controlled_2020}. 

In Fig.~\ref{sigma_all} we present the real and imaginary parts of the THz conductivity for all  films studied. The scattering rate $\Gamma = 1/2\pi\tau$,  which defines the frequency at which the real and imaginary parts of the conductivity intersect in the Drude model, is clearly visible within our spectral range upon cooling and for the cleaner samples the scattering rate moves to frequencies below our spectral range at the lowest temperatures. For high temperatures where $\omega \ll \Gamma $ is satisfied, the complex conductivity is observed to be largely frequency independent. At the lowest temperatures where $\omega \gg \Gamma$, we observe the sharp increase of the real and imaginary parts of the conductivity as the frequency decreases. The sharpness of the Drude peak observed at the lowest temperatures are indicative of the large mean free paths of our PdCoO$_2$ thin films. 

\begin{figure}[h]
    \includegraphics[width=1\columnwidth]{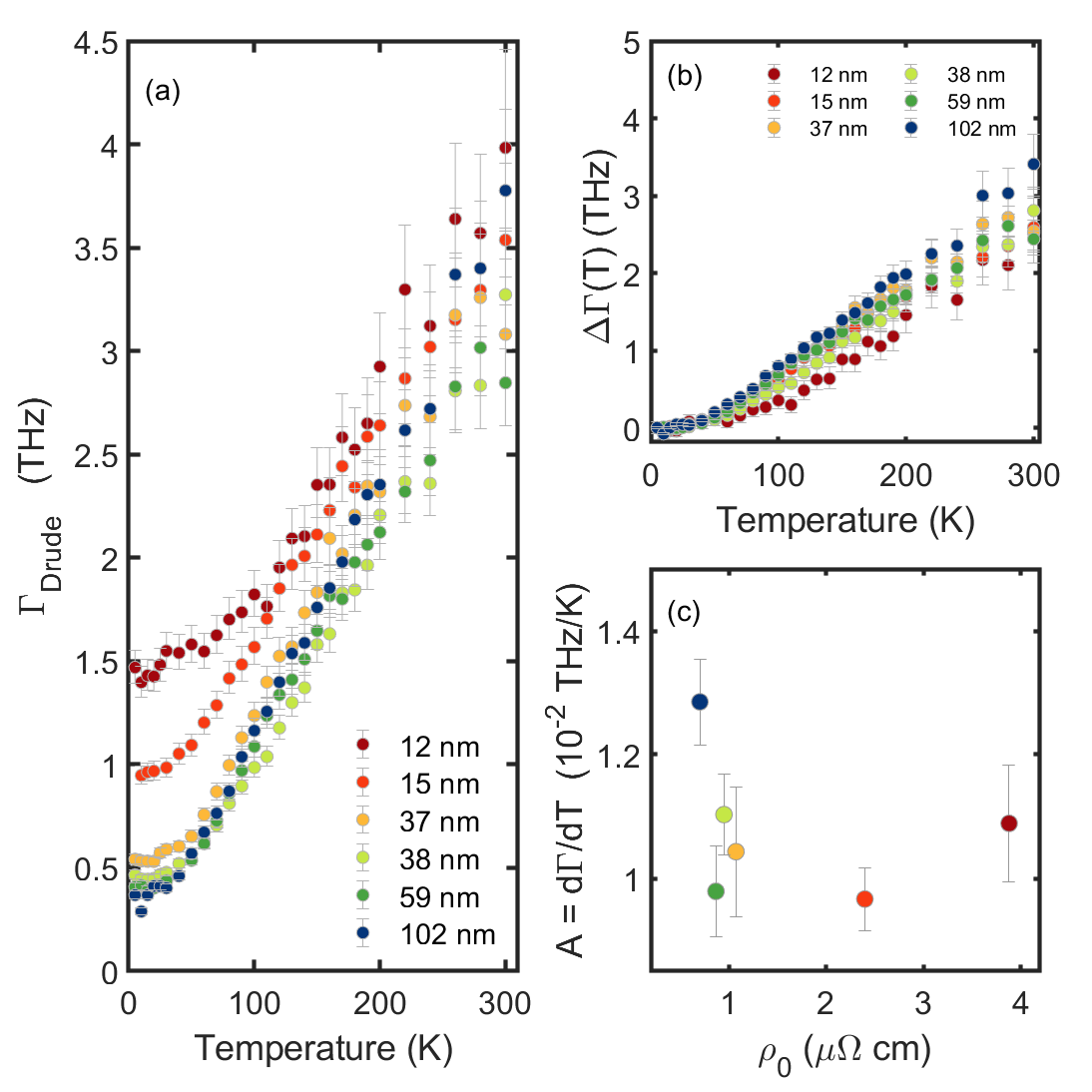}
    \caption{(a) The Drude scattering rate presented as a function of temperature. The error bars represent the change in fit parameters due to changes of $\pm 0.5 \;\mu$m of $\Delta L$ in the complex conductivity calculation from the THz transmission that reflect our experimental precision. The temperature dependence of the scattering rate given by $\Delta\Gamma(T) = \Gamma(T) - \Gamma(5 K)$ is shown in (b) for all samples to highlight systematic change in the scattering rate with disorder. The temperature dependence is parameterized for $T > 100$ K by the high temperature slope between 100-300 K and is shown in (c) with the residual resistivity $\rho_0$ of each sample. Note that the slope of the scattering rate has a much weaker dependence on film disorder compared to the temperature dependence exhibited by the resistivity in Fig.~\ref{DCResistivity}(b,c).}
    \label{ScatteringRate}
\end{figure}

As shown for selected temperatures in Fig.~\ref{sigma_35nm} (a,b) the complex conductivity fits very well to the Drude form.  Note that these fits are exceedingly well constrained by both the real and imaginary parts of the conductivity. In good agreement with previous Fourier-transform infrared spectroscopy (FTIR) measurements on single crystals \cite{homes_perfect_2019}, the large energy separation between the intraband and interband transitions as well as the absence of low frequency phonons support the excellent consistency with only a single Drude term in the THz spectral region.

From the Drude fits the temperature dependence of the fitting parameters e.g. the temperature dependent scattering rate $\Gamma$ and the spectral weight $\omega_p^2$, were obtained and are shown in Figs.~\ref{ScatteringRate} and ~\ref{Spectral Weight}. At the lowest temperatures for $T < 50$ K, the temperature dependence of the scattering rate in Fig.~\ref{ScatteringRate}(a) is small, consistent with what is observed in the dc resistivity. Matthiessen's rule would predict that the scattering rate for all samples should have the same temperature dependent behavior due to the identical scattering process in all films, with different residual values depending on the disorder level. When looking at the temperature dependence $\Delta\Gamma(T) = \Gamma(T) - \Gamma(5 K)$ of the scattering rate in Fig.~\ref{ScatteringRate}(b), in the low temperature regime we observe some differences in the temperature dependence between samples; in the high temperature regime the temperature dependence of the scattering rate is not large enough to explain the differences in resistivity.  In order to parameterize the high temperature dependence,  in Fig.~\ref{ScatteringRate}(c) we show the high temperature slope $d \Gamma/dT$ as a function of residual resistivity in the range of 100-300 K which appears to be very nearly independent of residual disorder.

Shifting focus to the Drude weight, $\omega_p^2$, as shown in Fig.~\ref{Spectral Weight}, we observe that all samples have very little temperature dependence; however, $\omega_p^2$ changes dramatically as a function of disorder. Comparing across our thickness range, we observe an unexpected 1.6$\times$ increase of the spectral weight from the most to the least disordered samples. This observation is surprising but it cannot be explained by random sample to sample variation of the carrier density ($\leq 20\%$), particularly because what is observed in the spectral weight $\omega_p^2$ has a strong systematic dependence on disorder. If one interprets the Hall coefficient as an inverse carrier density (Fig.~\ref{carrierDensity}), one finds that at low temperatures the 3D carrier density across the films is largely constant, $n = (2.15 \pm 0.07)\times 10^{22}$ cm$^{-3}$. While the carrier density in thin films is reduced from the bulk value of $n = 2.45\times 10^{22}$ cm $^{-3}$ \cite{takatsu_roles_2007}, there is no systematic disorder dependence when comparing within the films. When considering the mobility inferred from the Hall effect and the resistivity is $\mu = 50 - 200 \; \mathrm{cm}^2 \,\mathrm{V}^{-1} \mathrm{s}^{-1}$ at low temperature, indicating that even the thinnest films are still excellent conductors. Hence, variation of the carrier density with thickness cannot be the origin of the observed changes in spectral weight across the films studied.

\section{Discussion} 
\subsection{DMR in the High Temperature Regime}

\begin{figure}[t]
    \centering
    \includegraphics[width=1 \columnwidth]{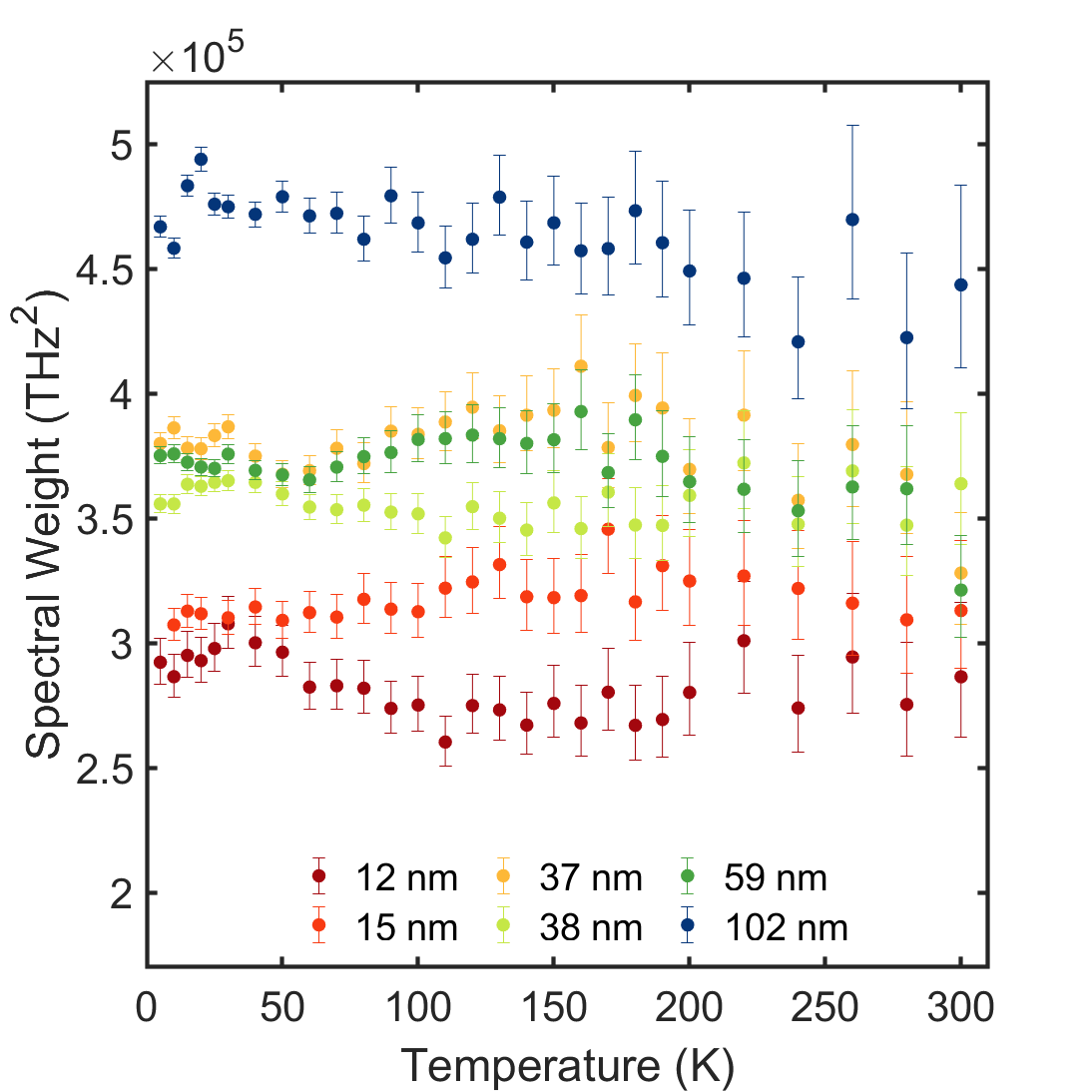}
    \caption{The THz spectral weight obtained from the Drude fits for all samples. The error bars represent the change in fit parameters due to changes of $\pm 0.5 \; \mu$m of $\Delta L$ in the complex conductivity calculation from the THz transmission. Error from the Drude fits is negligible.}
    \label{Spectral Weight}
\end{figure}

\begin{figure}[h]
    \centering
    \includegraphics[width=1 \columnwidth]{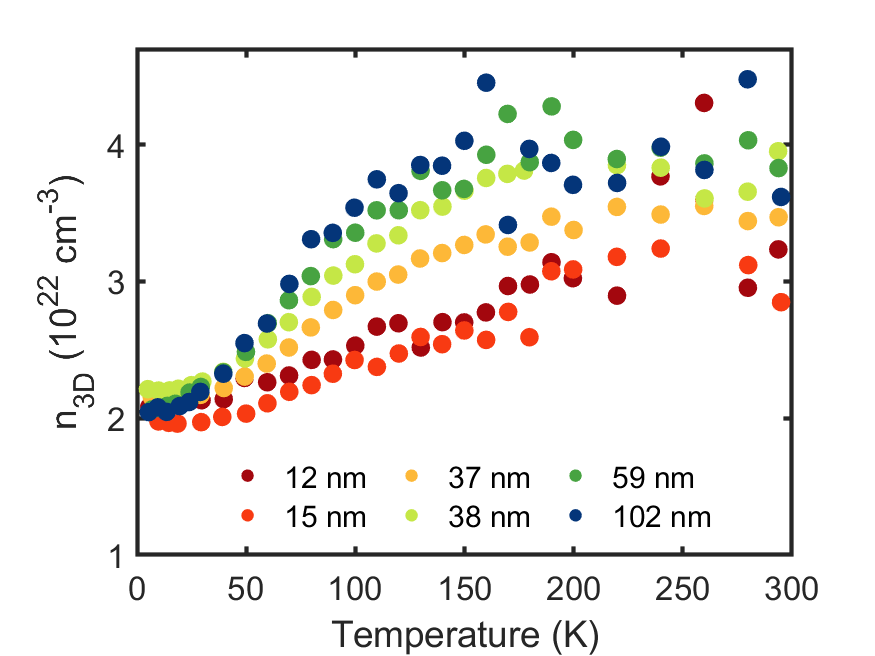}
    \caption{The temperature dependence of the carrier density as given by the Hall effect. While the low temperature carrier density for all films is comparable, the increase in carrier density at room temperature appears to be strongly dependent on the film thickness, with the thicker films approaching the factor of 2 increase observed in single crystals. }
    \label{carrierDensity}
\end{figure}

\begin{figure*}[!ht]
    \centering
    \includegraphics[width=1 \textwidth]{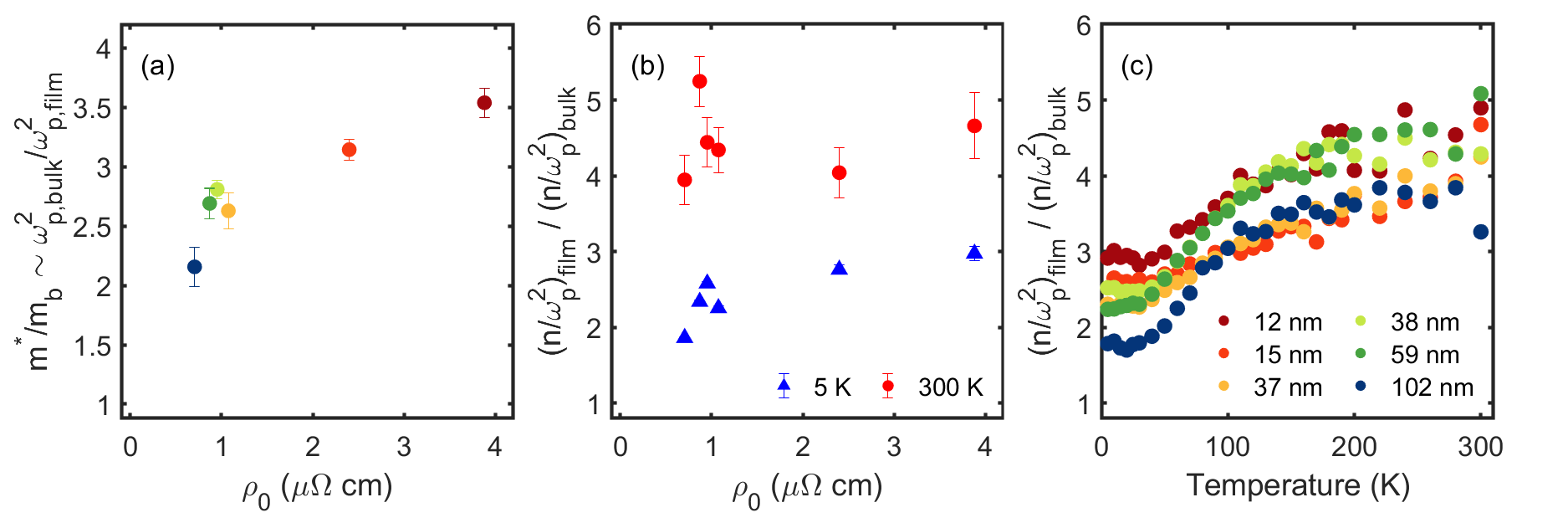}
    \caption{(a) Effective mass from ratio of Drude spectral weight of films referenced to that of single crystals. This ratio gives the effective mass $m^*/m_b$ = $\omega_{p, \textrm{bulk}}^2 / \omega_{p,\textrm{film}}^2$ from the renormalization of carriers near the Fermi energy in the films in an extended Drude picture. (b) Using the Drude spectral weight and the carrier density, the effective mass calculated $m^* \propto n/\omega_p^2$ at high and low temperatures are given. It is clear from both (a,b) that the effective mass increases with residual disorder. The normalization value used for the bulk is given by the bulk plasma frequency $\omega_p = 1000$ THz and low temperature carrier density $n_{3D} = 2.45\times10^{22}$ cm$^{-3}$ \cite{homes_perfect_2019,hicks_quantum_2012,mackenzie_properties_2017}. In (b), it is clear that the effective mass increases with both disorder and temperature, with more disorder dependence evident at low temperatures. In (c) the same normalized effective mass is shown as a function of temperature. It is clear the the effective mass increases as the film disorder increases, but all samples also show a similar upwards trend for the effective mass as a function of temperature.  The effective mass ratio appears saturate as the temperature approaches 300 K. }
    \label{effective mass}
\end{figure*}

\begin{figure}[h]
    \centering
    \includegraphics[width=1 \columnwidth]{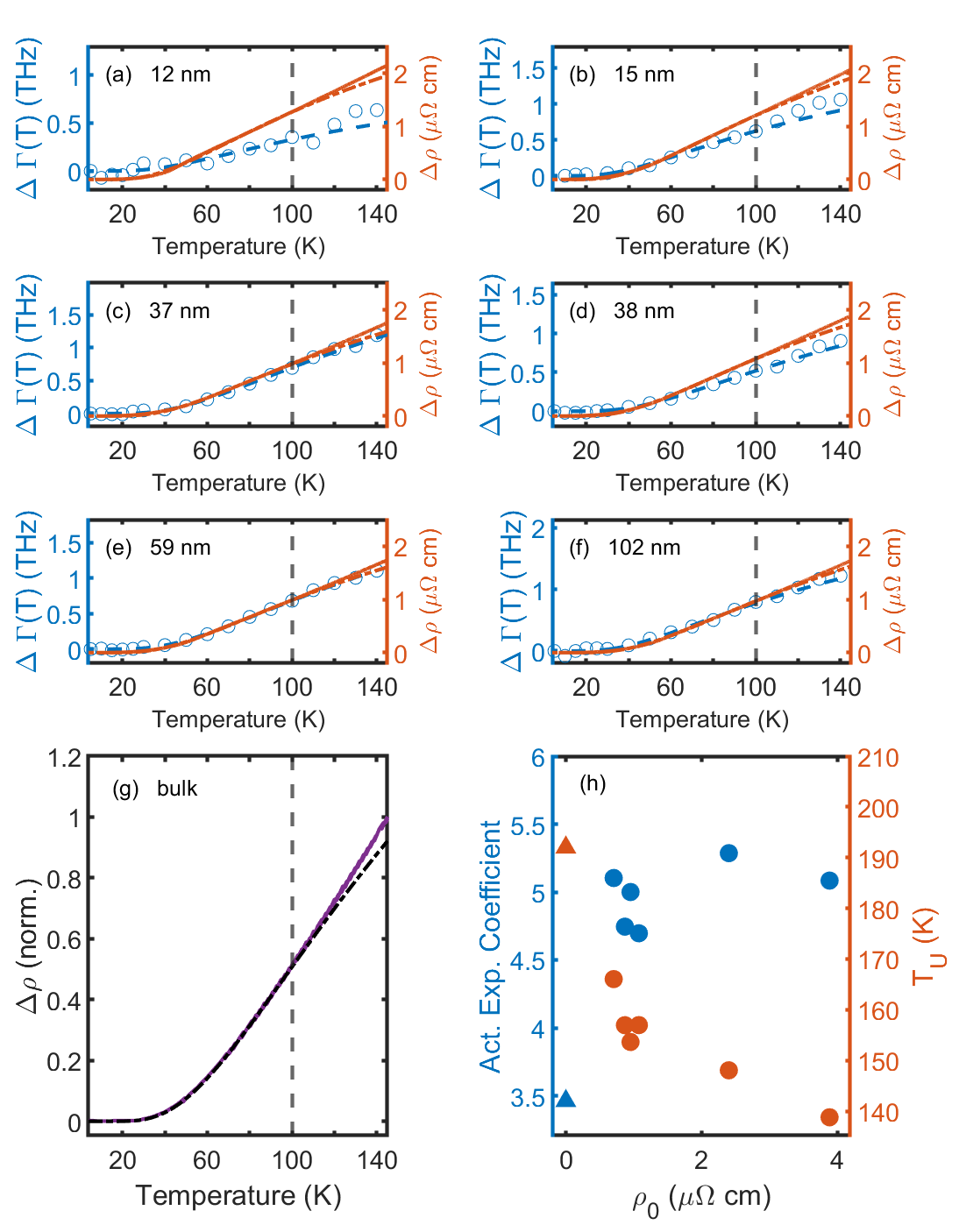}
    \caption{Two parameter fitting results for the activated exponential form $\Gamma = A \textrm{ exp}(-T_U/T)$. The dashed vertical line demonstrates the 100 K temperature cutoff used for the low temperature fitting. We find the in subplots (a-f) both the THz scattering rate (points) and the dc resistivity data (solid line) are shown; note the good consistency of the fits to the experimental data. For comparison, the dc resistivity of a bulk sample is shown in the bottom left (g). We observe good consistency to the activated exponential form for all samples in both the THz scattering rate and resistivity. The fitting coefficients are given as a function of sample disorder in (h). Note that the coefficient of the Umklapp scattering term increases with sample disorder and the characteristic Umklapp $T_U$ energy term decreases with disorder. This is suggestive of disorder-mediated Umklapp scattering that relaxes the restrictive momentum and energy conservation cases necessary in the case of very pure crystals. }
    \label{actExponentialFit}
\end{figure}

In order to better understand the disorder dependence of the resistivity, we first address the issue of the DMR in the high temperature regime. In general, for high temperatures $T > \Theta_D/5$, the electron-phonon contribution to resistivity or the scattering rate should have a T-linear dependence. In the earliest work on high quality crystals, the in-plane resistivity was found to demonstrate a super-linear trend with $\rho_{ab} \propto T^{1.4}$ between 100 and 300 K \cite{takatsu_roles_2007}. This was found to be reproduced in PLD grown thin films (8-15 nm) across a temperature range of 50 - 300 K, albeit with a much larger residual resistivity. This mechanism responsible for the deviation from simple T-linear behavior was identified due to scattering from high frequency optical phonons \cite{takatsu_roles_2007}. This nontrivial coupling of the electron fluid to high frequency phonons has also been observed in ReO$_3$ \cite{allen_bloch-boltzmann_1993} and MgB$_2$ \cite{masui_phonon_2002,masui_normal_2003}. One alternative contribution to resistivity that would give deviations from T-linear behavior would be electron-electron scattering $\rho_{ee} \propto T^2$; however, the temperature dependence from our work and previous works have consistently shown that the resistivity in PdCoO$_2$ is inconsistent with Fermi-liquid like scattering with $\rho \propto T^2$ at low temperatures. While electron-electron scattering can also give rise to deviations from T-linear resistivity at low and intermediate temperatures, in PdCoO$_2$ the small electronic specific heat coefficient $\gamma = 1.28 \textrm{mJ/ mol K}^2$ and large carrier densities ($n \sim 10^{22} $ cm$^{-3}$) suggest that electron-electron interactions are unlikely to be the dominant contribution to resistivity as $\rho_{ee} \propto (1/n)^2$ \cite{takatsu_roles_2007}. The absence of clear electron-electron scattering in the resistivity also precludes more exotic behaviors reliant on strong electron-electron interactions such as hydrodynamic effects or inter-layer Coulomb drag\cite{liao_drag_2020}.

For the films considered in our work, we find that the resistivity and scattering rate at high temperatures $T > 100$ K are better described by T-linear behavior rather than power-law $T^{1.4}$ behavior observed previously. For the purpose our study, we use the slope of the T-linear resistivity to parameterize the temperature dependence of the resistivity in the high temperature regime to explore the origins of the observed DMR (T $> 100 $K). The observation of a high-temperature DMR in delafossites has only been explored in electron irradiation   on PtCoO$_2$ and PdCoO$_2$ single crystals \cite{sunko_controlled_2020}, where the addition of disorder leads to a change in the relevant scattering mechanisms. However, the range of additional disorder induced by irradiation ($\sim 0.5 \,\mu\Omega \,\mathrm{cm}$) is much smaller than the range of disorder accessible in thin films by tuning annealing and thickness ($\sim 3 \, \mu\Omega\, \mathrm{cm}$).  We would note that a similar increase of the temperature dependent component of the resistivity with increasing residual resistivity has been previously observed in extruded thin films of aluminum \cite{van_zytveld_size-dependent_1969}. \\

The most relevant scattering mechanism in the high temperature regime is electron-phonon scattering, which demonstrates increasing DMR with increasing disorder. Yet the temperature dependent inelastic component of the scattering rate appears to be only weakly dependent on the elastic scattering from residual disorder at high temperatures as seen in Fig.~\ref{ScatteringRate}(b,c). Counter intuitively, we instead see a strong dependence of the spectral weight on residual disorder as shown in Fig.~\ref{Spectral Weight}. This suggests that the DMR is driven not by changes in the scattering rate but rather in changes in the spectral weight, where the dc resistivity is given by the ratio of the two quantities $\rho(T) = (1/ \omega_p^2(T)) \times (1/\tau(T))$. Interpreting the Hall effect measurements in terms of carrier density, it appears that the carrier density at low temperatures changes less than 5\% between all samples, well within expected small sample to sample variation. Within this interpretation, the systematic change in the spectral weight as a function of disorder should be thought of as an enhanced effective mass arising in these thin films. 

Another method to parameterize the effective mass is to combine the Hall effect measurement with the optical spectral weight. While the Hall effect measurement shows strong temperature dependence as seen in Fig.~\ref{carrierDensity} with a change of 1.5-2x between low and high temperature, this change in consistent with what is observed in bulk samples \cite{nandi_unconventional_2018}. The temperature dependence in the bulk was interpreted as violation's of Kohler's rule due to k-dependent scattering. If we interpret the Hall effect measurement as the carrier density, then the effective mass in this material must have the same temperature dependence of the carrier density since $\omega_p^2 \propto n/m^*$.

While disorder does not change the magnitude of the carrier density in the low temperature regime, the increase in the carrier density at high temperatures is strongly dependent on the film thickness. This implies that the effective mass in these films is both temperature and disorder dependent. This is presented in Fig.~\ref{effective mass}, where a normalization factor from the single crystal carrier density and plasma frequency at 5 K is used \cite{homes_perfect_2019,takatsu_roles_2007}. Regardless of whether one calculates an extended Drude type effective mass $m^*_{EDM} = \omega_{p,bulk}^2 / \omega_{p,film}^2$ in Fig.~\ref{effective mass}(a) or the effective mass calculated from the ratio of the carrier density  and spectral weight  $m^* = ne^2/\omega_p^2$ seen in Fig.~\ref{effective mass}(b,c), there is clear evidence that the quasiparticle mass is enhanced by at least a factor of 2 over single crystals using either metric.  One advantage of the latter measure of the effective mass from the carrier density from the Hall effect and spectral weight shown in Fig. \ref{effective mass}(b,c) is that the ratio is independent of any errors in the experimentally measured thickness of the film. In addition to the clear relationship between mass enhancement and disorder, the temperature dependence of the effective masses are particularly interesting, as all films show an effective mass that increases with temperature that appears to saturate as the temperature approaches 300 K.  One caveat is that the temperature dependent effective mass proportional to the Hall effect measurement, which may be in a regime where it cannot be directly interpreted as the actual carrier density. Yet, regardless of whether one uses an extended Drude effective mass or the effective mass from the Drude formula, it is clear that the mass is enhanced significantly as a function of disorder in these PdCoO$_2$ thin films. While the exact mechanism is not obvious, we will discuss different scenarios consistent with our experimental observations. 

\subsection{Disorder Enhancement of the Electron-Phonon Interaction}
The most basic mechanism for an effective mass arising from the electron-phonon coupling has been explored in the Eliashberg-McMillan theory, where the electron-phonon coupling constant $\lambda = 2 \int_0^{\infty} \frac{d\omega}{\omega} \alpha(\omega)^2F(\omega)$ is related to the Eliashberg-McMillan function $\alpha(\omega)$ and the phonon density of states $F(\omega)$. In the simplest picture, one can relate the electron-phonon coupling constant to the dc resistivity at high temperatures via $\rho = (2\pi m k_B/ne^2 \hbar) \,\lambda_{tr}  \,T$, where $m$ is the mass, $n$ is the carrier density, $T$ is the temperature and $\lambda_{tr}$ is the transport coupling constant, which differs from the intrinsic value by the $(1-\textrm{cos}\theta)$ forward scattering suppression factor \cite{grimvall_electron-phonon_1976}. However, the mass enhancement due to electron-phonon coupling  $m^* \sim (1+\lambda)$ is usually small since for noble metals $\lambda \sim 0.1$ \cite{lin_electron-phonon_2008}, but can become enhanced $\lambda \sim 1 - 3$ for heavy elements like Pb, Bi or Ga \cite{grimvall_electron-phonon_1976}. Yet, in single crystals of PdCoO$_2$ the electron-phonon couplings is $\lambda < 0.05$, much less than other noble metals \cite{takatsu_roles_2007,hicks_quantum_2012}. While the relative difference of the slope of the resistivity at high temperatures is small between films, it is unlikely that the slope of the resistivity (and hence coupling constant) can increase by 50\% going from a bulk sample to a thin film without major changes in the underlying scattering mechanism as seen in our THz scattering rate. Hence, a disorder dependent change in the intrinsic electron-phonon coupling constant $\lambda$ is cannot be the primary explanation for the observations of DMR in the resistivity in Fig.~\ref{DCResistivity}(c) and the enhanced effective mass in  Fig.~\ref{effective mass}. 

Other experimental work have found evidence for DMR and disorder-enhanced behavior beyond simple electron-phonon interactions. In recent work on epitaxial films of Cu, the electron-phonon coupling was found to increase for more disordered films and exhibited a temperature dependent DMR $\Delta\rho(T)$ that increases with disorder \cite{timalsina_evidence_2013}. By combining dc resistivity and optical pump probe experiments, the slope of the resistivity at high temperatures, proportional to the electron-phonon coupling $\lambda$, was found to increase with decreasing film thickness.  

The interaction of scattering mechanisms in the dirty limit for metals has been previously studied for a variety of metallic thin films; for sufficient levels of disorder, an additional inelastic electron scattering channel from vibrating impurities can arise in tandem with the intrinsic electron-phonon scattering mechanism \cite{ptitsina_electron-phonon_1997}. This additional electron-phonon-impurity interference scattering channel $\frac{\Delta \rho}{\rho_0} \sim T^2$ appears as an additional term in the resistivity for $T < \Theta_D$ and should be the dominant contribution over the Bloch-Gruneisen term in the low temperature regime \cite{ilin_interrelation_1998}. This was experimentally found to be the case for ion-irradiated NbC thin films albeit for much higher residual resistivities (16-98 $\mu\Omega$ cm) as well as the earlier study which studied elemental Al, Be, Nb and Au films \cite{ptitsina_electron-phonon_1997}. The interaction between electrons and transverse phonons are thought to be the dominant driver of this interaction. While the resistivity of our thin films of PdCoO$_2$ does not follow the same $T^2$ dependence expected at low temperatures, it is clear that the mechanism behind our observed DMR must come from correlations between scattering mechanisms. 

Other possibilities include phonon localization due to disorder as found in graphene devices \cite{evangeli_experimental_2021} or from modifications of the phonon density of states which enhance the electron-phonon scattering channels \cite{berry_deviations_1972}. These studies show that additional mechanisms beyond simple electron-phonon scattering seem likely candidates for explaining the temperature dependence of the DMR observed and the enhancement of the effective mass over bulk samples. This does not rule out the possibility of more exotic effects such as strong electron-phonon coupling, a coupled polaron fluid or hydrodynamics that could be responsible for the behavior observed, albeit through the mediation of strong disorder in the films studied. \\

\subsection{Low Temperature Enhancement of Activated Exponential Scattering}
Turning our attention to the low temperature regime $T < 100$ K, it appears that the resistivity and THz scattering rate $\Gamma (T)$ are both well-described by the activated exponential form $A\; \textrm{exp}(-T_U/T)$ as shown in Fig.~\ref{actExponentialFit}(a-f). This form for the low temperature resistivity was first used by Hicks \emph{et. al.} \cite{hicks_quantum_2012} to model the low temperature resistivity in PdCoO$_2$, where they found $T_U = 165$ K for in plane transport using $T = 30$ K as the cutoff temperature for fitting.  The exponential form arises from the fact that for Fermi surface electrons that do count cross the momentum space boundary for Umklapp scattering must be thermally activated to an energy where they can so scatter.  From our resistivity measurements and the THz scattering rate, we find that the data is in good agreement with activated exponential form using a two parameter fit. However, due to our lower sensitivity to the small changes in resistivity at small temperatures, we find that the activated exponential form can be well fit using a maximum temperature cutoff of 60 - 100 K. This is indeed measured up to much higher temperature range with respect to the original \cite{hicks_quantum_2012} discovery of the activated exponential form for $T < 30 $ K with $T_U = 165$ K.   More recently grown single crystals do display good agreement to the activated exponential form up to $T = 100$ K \cite{li_situ_2019} as shown in Fig.~\ref{actExponentialFit}(g) with a slightly higher $T_U = 190$ K. The activated exponential prefactor $A$ and the Umklapp temperature $T_U$ are shown in Fig.~\ref{actExponentialFit}(h) using a high temperature cutoff of 100 K. We observe that the prefactor $A$ increases with disorder while the Umklapp temperature $T_K$ is a decreasing function of residual disorder. This is consistent with the observation of smaller temperature dependence in the scattering rate for more disordered films as seen in Fig.~\ref{actExponentialFit}(a-f) as parameterized by the lower activated exponential prefactor and the lower characteristic Umklapp temperature $T_U$. 

While a clear physical mechanism for this characteristic dependence is not obvious, we can rule out effects of disorder directly changing the energy gap associated with Umklapp scattering due to the distance between adjacent Brillouin zones, since the carrier densities are not expected to change dramatically with disorder. Rather, the dependence of the Umklapp temperature on disorder suggests that disorder may play a role in reducing the strict momentum conservation conditions necessary for Umklapp scattering, resulting in a lower effective $T_U$. While an activated exponential behavior of the resistivity due to Umklapp scattering has been previously observed in the alkali metals \cite{ziman_electrical_1954,bailyn_transport_1958,bailyn_transport_1960}, the characteristic temperature scale ($T_U \sim 20$ K) and onset temperature ($T \leq 4 K$) are significantly lower than what is found in PdCoO$_2$ \cite{bass_temperature-dependent_1990}.

One open question is how closely associated this activated exponential form is to the presence of phonon drag, where the dominant contribution to resistivity from Umklapp electron-phonon scattering process at low temperatures indicate that inability of normal electron-phonon scattering processes to relax momentum. At low temperatures, there is clear evidence of a phonon-drag peak in PdCoO$_2$ in the thermopower in single crystals that peaks around $\Theta_D/5 \approx  60 $ K \cite{daou_large_2015}. However, the presence of the phonon drag peak is sensitive to disorder, and is not observed in PLD grown films \cite{yordanov_large_2019}. Due to the lack of thermopower measurements on MBE films, it is unclear whether the activated exponential form can be independent of signatures of phonon drag.

\section{Conclusion} 

In summary, from the dc resistivity and the complex THz conductivity, we believe we observe a disorder-enhanced electron-phonon contribution to the dc resistivity in PdCoO$_2$ thin films that scales with residual disorder levels. The films studied show clear deviations from Matthiessen's rule present in the dc resistivity , where the temperature dependent component increases in magnitude with increasing disorder. This suggests that the inelastic scattering channels dominated by electron-phonon mechanisms are not fully independent of disorder levels across the samples. To further examine the origin of the DMR, we find that from fitting the THz conductivity that the scattering rates are almost independent of the residual disorder at high temperatures. Rather, we find that the Drude spectral weight is an decreasing function of disorder, manifesting in an effective mass enhancement of $\geq 2$ over the bulk value. This finding is highly anomalous and robust using both the ratio of the spectral weight and combining the Drude spectral weight and the carrier density from Hall effect measurements. This suggests that the DMR observed in the high temperature resistivity is due to additional electron-phonon scattering pathways mediated by disorder.

Furthermore, we find that in the low temperature regime the resistivity and THz scattering rate are well described by the activated exponential form due to Umklapp electron-phonon scattering when both normal electron-phonon processes are frozen out at low temperature and Umklapp electron-phonon processes have an associated energy due to closed Fermi surfaces that do not intersect the boundaries for Umklapp scattering. We find a systematic disorder dependence of the temperature dependence at the low temperatures, where the characteristic Umklapp temperature $T_U$ is reduced for more disordered samples. This observation further demonstrates the effect of disorder-enhanced scattering mechanisms relevant for transport in PdCoO$_2$. We find that same temperature dependence of the resistivity in extremely pure single crystals also found in thin films.  Moreover, film thickness can be used to effectively tune the regime of transport in this class of materials. The evidence for large mass enhancement in these thin films remains an intriguing observation that will require further study. In particular, further transport experiments to characterize the \textbf{k}-dependence of scattering mechanisms such as magnetoresistance and thermopower measurements will be useful to elucidate how electronic scattering is modified in thin films of PdCoO$_2$.

\section{Acknowledgments} 
This work was supported by Army Research Office, Grant No. W911NF2010108 and the NSF DMR2004125.  Further support at JHU was supported by the Gordon and Betty Moore Foundation, EPiQS initiative, Grant No. GBMF-9454. 

DB and AL conducted the THz measurements. GR grew and characterized the films, and conducted the dc transport measurements, and SO supervised. DB and AL carried out the analysis. All authors contributed to the final manuscript.

\bibliography{PdCoO2_references}

\end{document}